# *Self-Consistent C-V Characterization of Depletion Mode Buried Channel InGaAs/InAs Quantum Well FET Incorporating Strain Effects*

Imtiaz Ahmed[1,2], Iftikhar Ahmad Niaz[*1,2], Md. Hasibul Alam[1]
[1]Department of Electrical and Electronic Engineering
[1]Bangladesh Univ. of Eng. & Tech.
[1]Dhaka, Bangladesh
[*]iftikhar.oni@gmail.com

Nadim Chowdhury[1], Zubair Al Azim[1,2], Quazi Deen Mohd Khosru[1,2]
[2]Department of Electrical and Electronic Engineering
[2]Green Univ. of Bangladesh
[2]Dhaka, Bangladesh

***Abstract*—We investigated Capacitance-Voltage (C-V) characteristics of the Depletion Mode Buried Channel InGaAs/InAs Quantum Well FET by using Self-Consistent method incorporating Quantum Mechanical (QM) effects. Though the experimental results of C-V for enhancement type device is available in recent literature, a complete characterization of electrostatic property of depletion type Buried Channel Quantum Well FET (QWFET) structure is yet to be done. C-V characteristics of the device is studied with the variation of three important process parameters: Indium (In) composition, gate dielectric and oxide thickness. We observed that inversion capacitance and ballistic current tend to increase with the increase in Indium (In) content in InGaAs barrier layer.**

## I. INTRODUCTION

Alternate materials and alternate structures have attained the focus of the researchers as silicon based CMOS has reached its fundamental limit [1]. Alternate structures like silicon based multi-gate devices like double gate [2], tri-gate [3], gate-all-around (GAA) [4] transistors are fabricated and systematically studied. These novel transistors can effectively improve short channel effects. Besides that research in alternative semiconductor material has emerged with the potential of improving drive current by finding a suitable high mobility material system. III-V materials have almost 100×higher electron mobility than Si [5]. So devices fabricated with III-V material can provide higher drive current. Buried channel high electron mobility transistors (HEMT) have also got the focus of the research community in recent years [6] as it is a promising device based on III-V.

This type of transistors has the advantage of avoiding carrier scattering due to the use of intrinsic material as the channel of the device. Both schottky barrier gate and oxide barrier gate has been reported for these devices [7]-[8] fabricated with nitride materials. Since the growth of nitride material is difficult, recently some HEMTs are reported to be fabricated using InAs and/or GaAs material [9]. Among these, Buried Channel InAs MOSFET is notable due to the fact that it has oxide gate which will reduce the gate leakage current [10]. Self-consistent analysis is yet to be done for this device to investigate the effect of material composition and gate dielectric variation on device performance.

## II. SIMULATION METHODOLOGY

We analyzed C-V characteristics and ballistic transport performance of the structure in Fig. 1. We obtained the results for the device solving 1D coupled Poisson and Schr□dinger equations [11]. The ballistic current was calculated with the model shown in [12]. We also incorporated strain in our simulator [13]. The band diagram of the device along with carrier profile at zero bias for two different In compositoins at top barrier region is shown in Fig. 2.

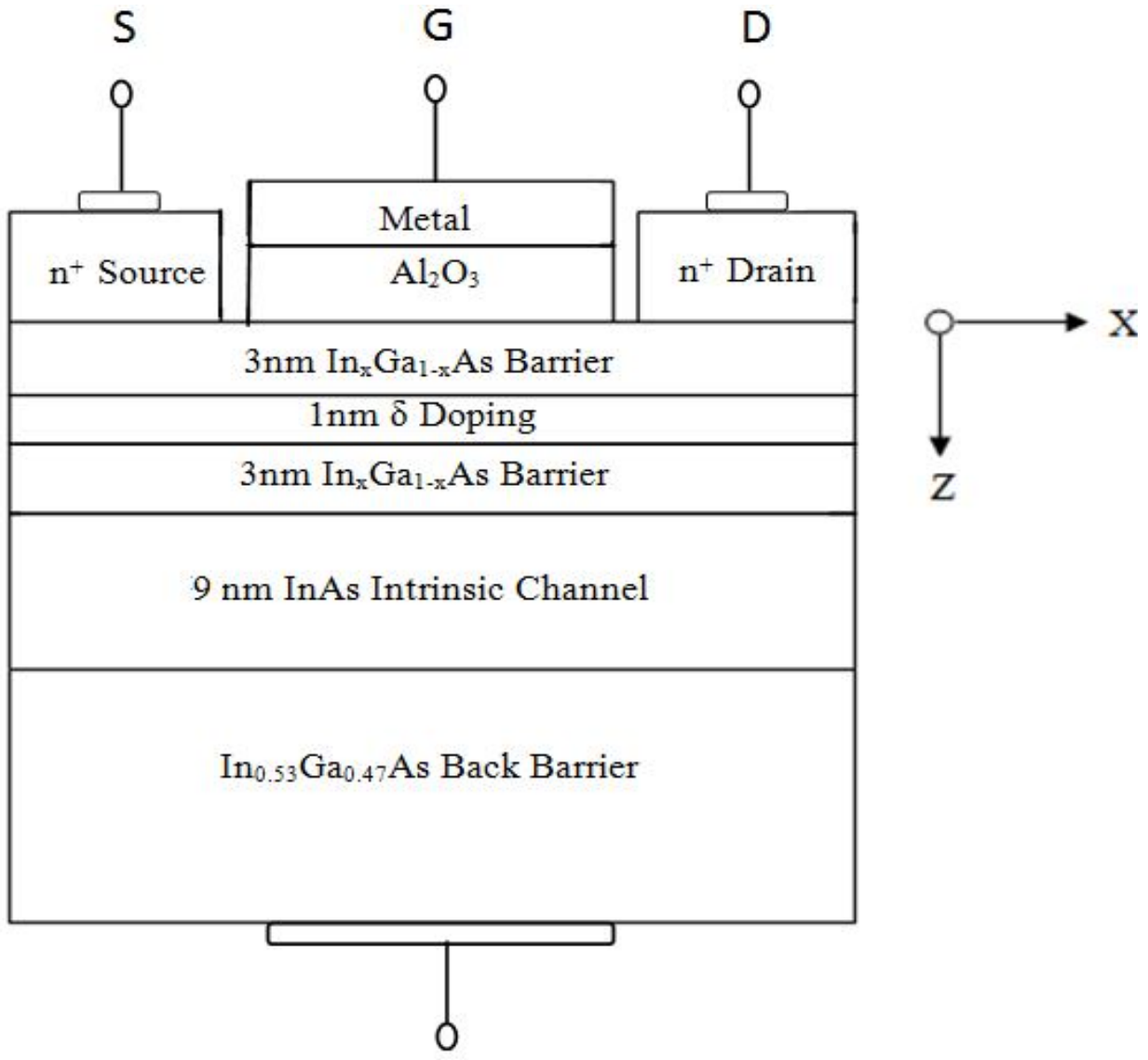

Figure 1. Basic device structure of Buried Channel InGaAs/InAs QWFET.

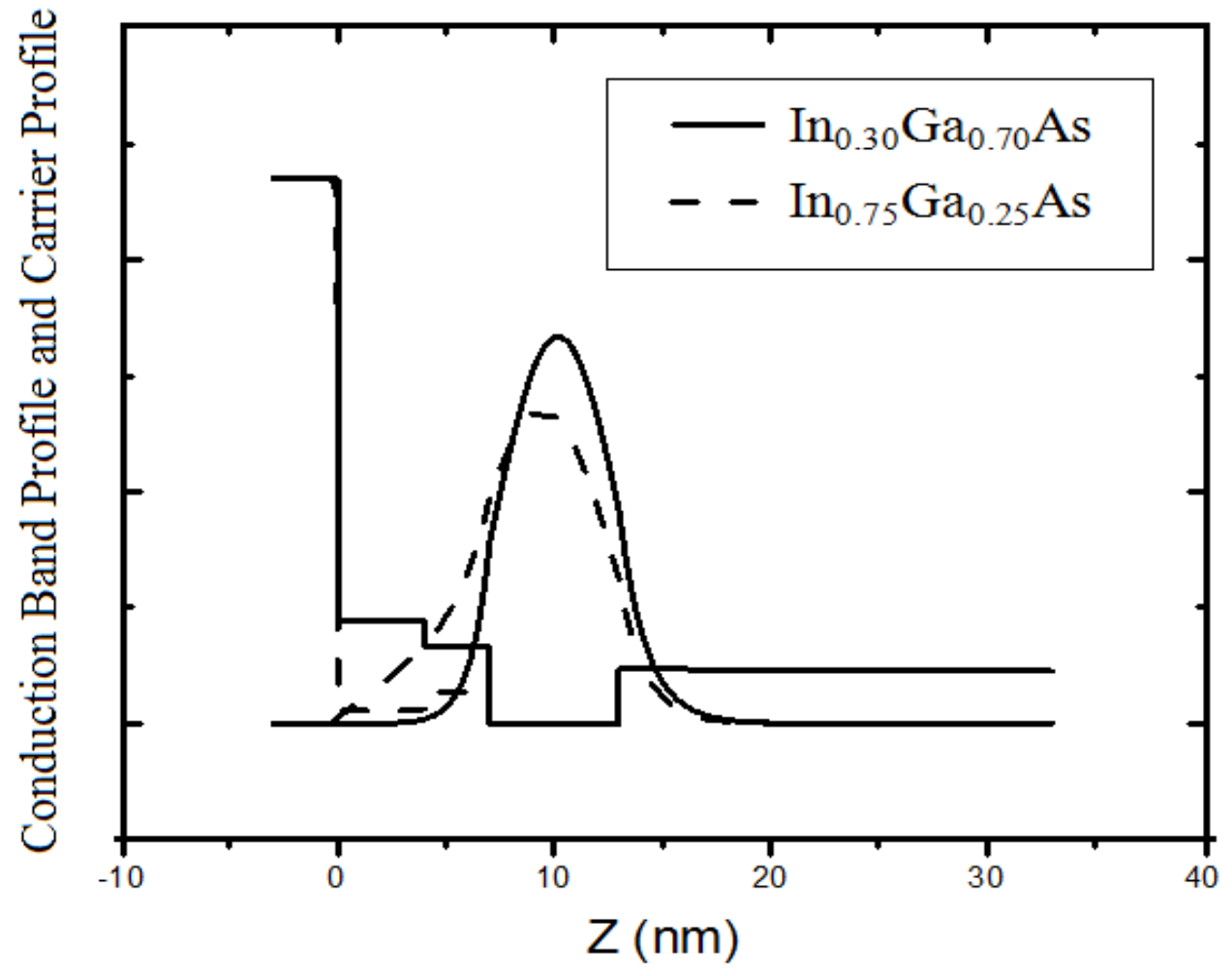

Figure 2. Conduction band profile along with carrier profile for two In compositions (not drawn to scale)

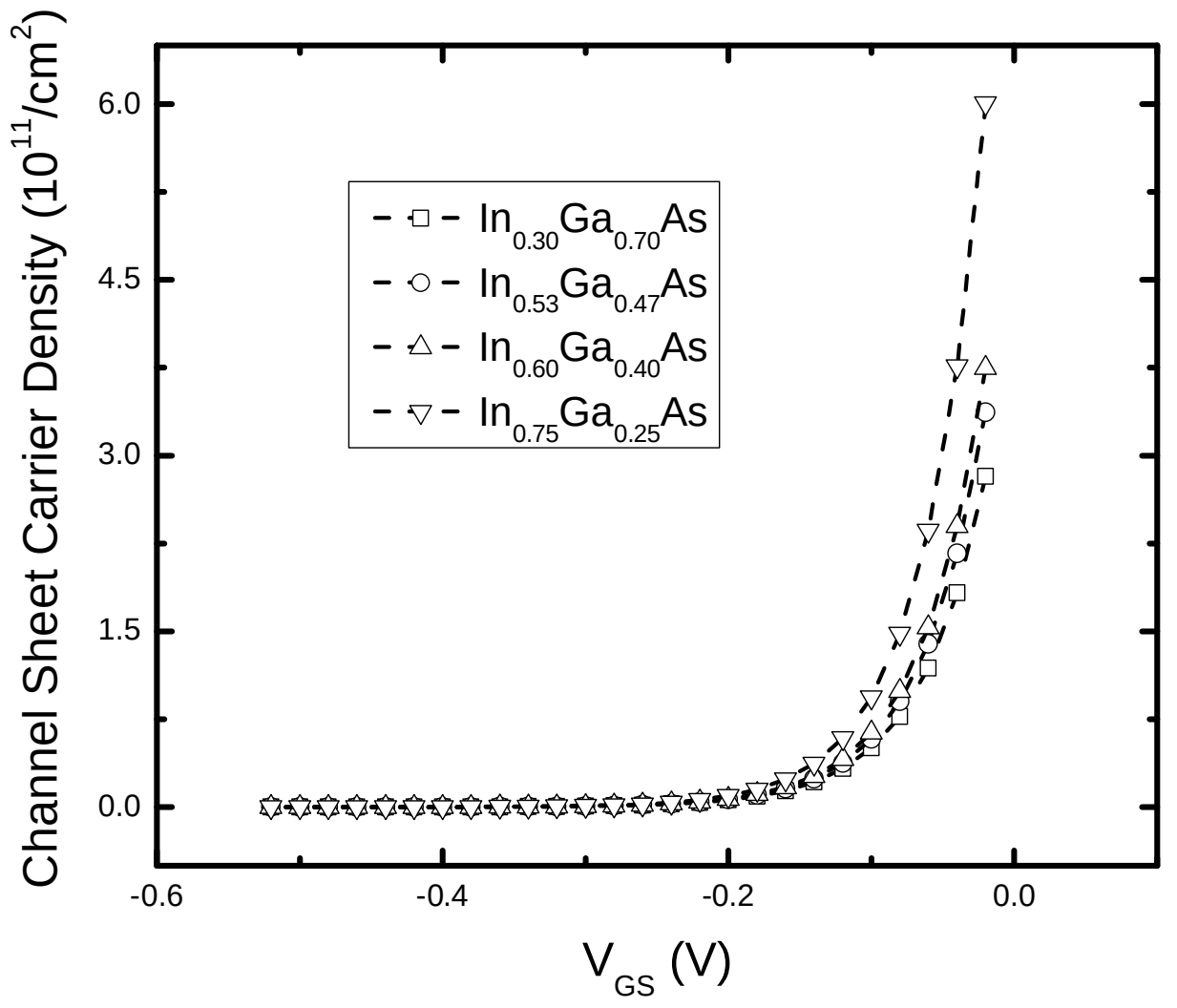

Figure 4. Channel sheet carrier density as a function of gate voltage for different In compositions.

## III. Results And Discussions

Eigen state denotes the subband minima for different subbands in the quantum well of the device formed due to the applied gate voltage. Fig. 3 presents the first eigen state of the device as a function of gate voltage for four different In compositions and hence different amount of strain between top barrier and channel. The eigen energy decreases with increasing In composition. With higher In percentage the band offset between top barrier and intrinsic channel layer decreseas which results in the lowest eigen energy with highest In composition and hence higher compressive strain.

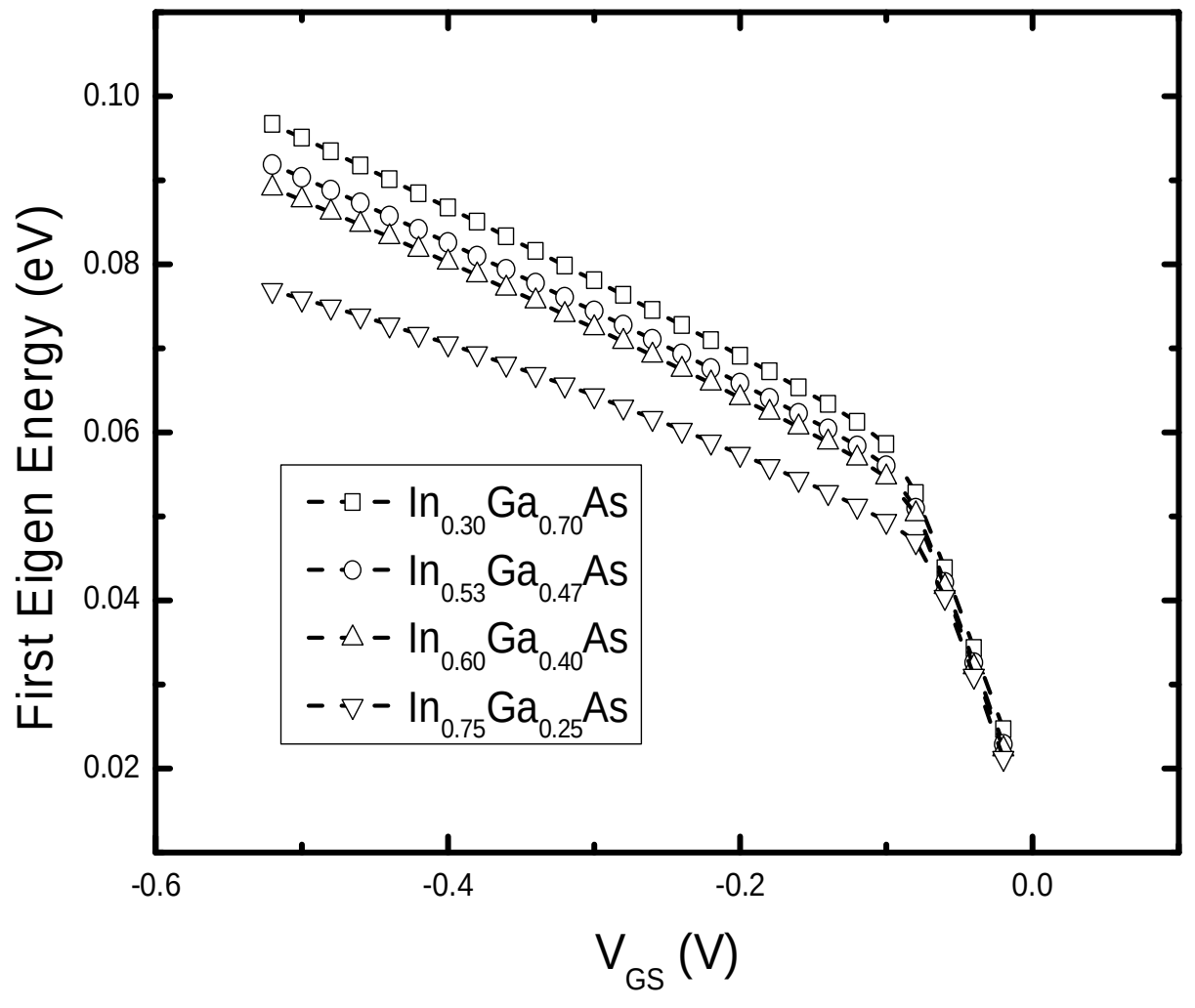

Figure 3. First eigen energy as a function of gate voltage for different In compositions.

Fig. 4 shows the channel sheet carrier density for different In compositions. The higher the In composition i.e the higher the tensile strain the higher the sheet carrier density. This can be explained from decreased eigen energy which accomodates more carriers in the potential well. The gate capacitances of the device for four different compositions is illustrated in fig. 5. It is observed that the capacitance is higher for higher In percentage which can be easily deducted from the higher sheet carrier density at channel region for higher tensile strain.

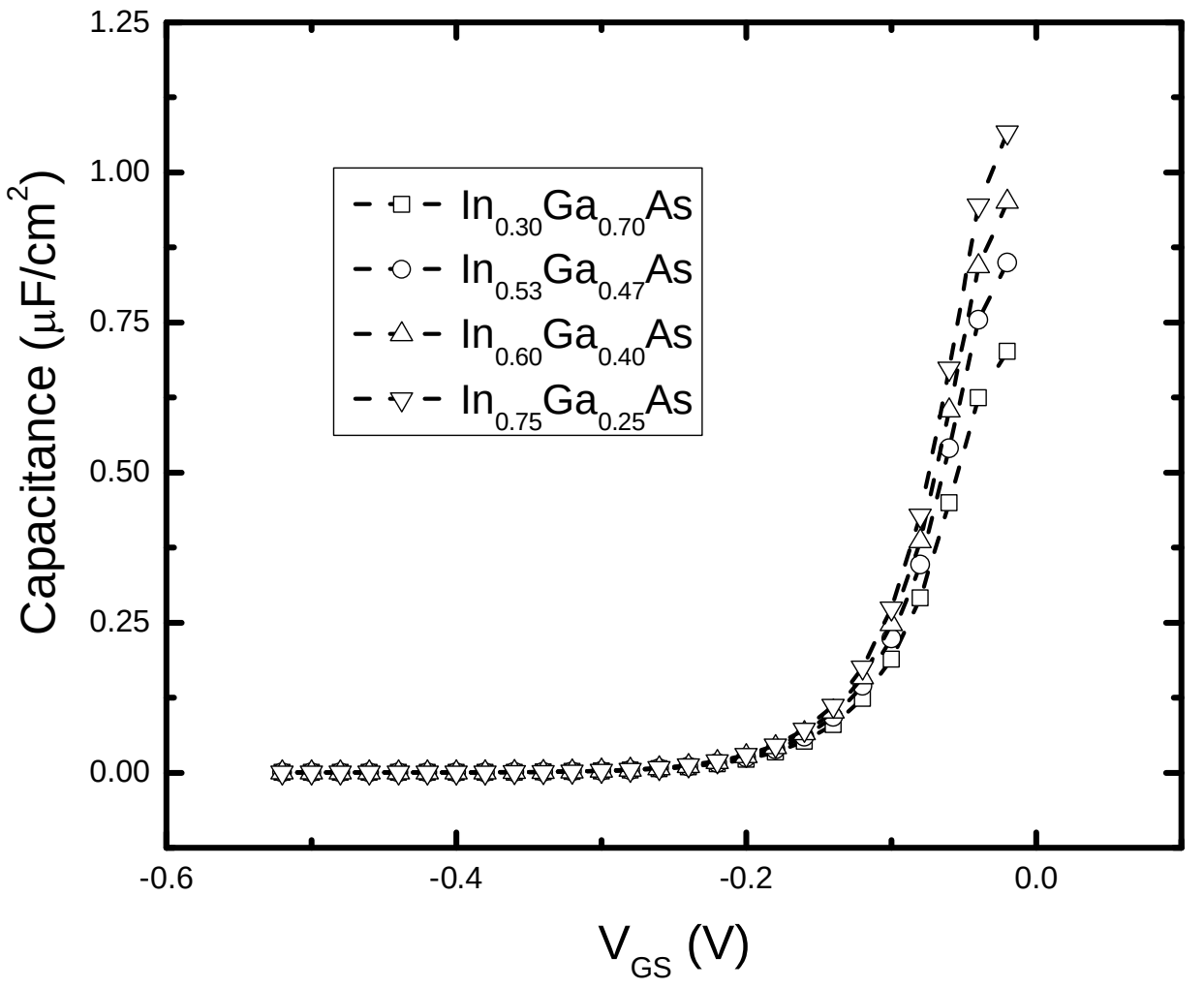

Figure 5. Gate capacitance as a function of gate voltage for different In compositions.

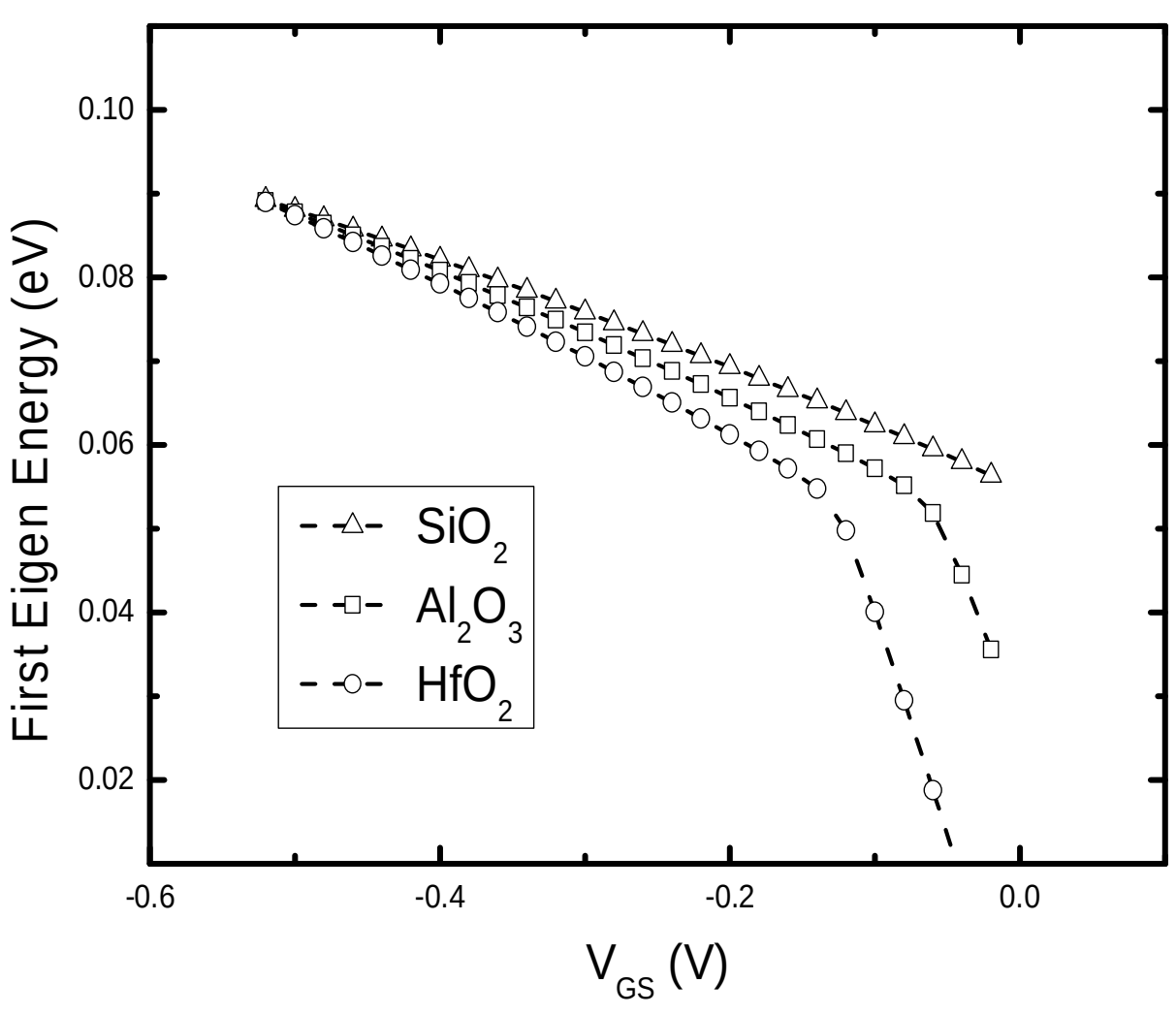

Figure 6. First eigen energy as a function of gate voltage for different dielectric materials for $t_{ox}$= 4 nm and In percentage 0.60

It is apparent that for the same physical oxide thickness but less Equivalent Oxide Thickness (EOT) in case of high-κ dielectrics the first eigen energy is decreasing. Channel sheet carrier density is highest for $SiO_2$ and lowest for $HfO_2$ before the cross over point after which the pattern reverses. The same reason explains the highest capacitance for $SiO_2$ before the cross over and the highest capacitance for $HfO_2$ after the cross over in Fig. 8.

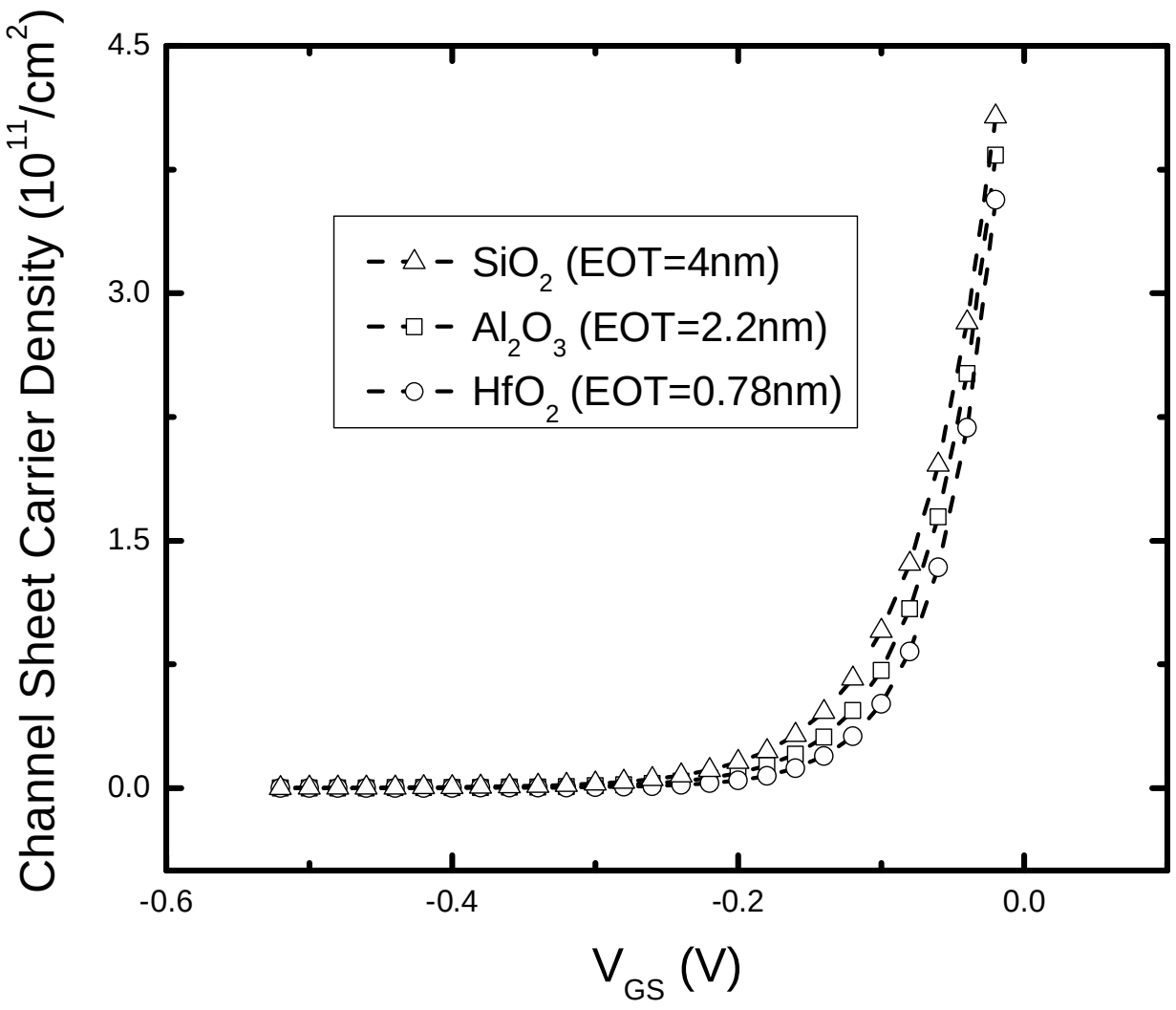

Figure 7. Channel carrier sheet density as a function of gate voltage for different dielectric materials.

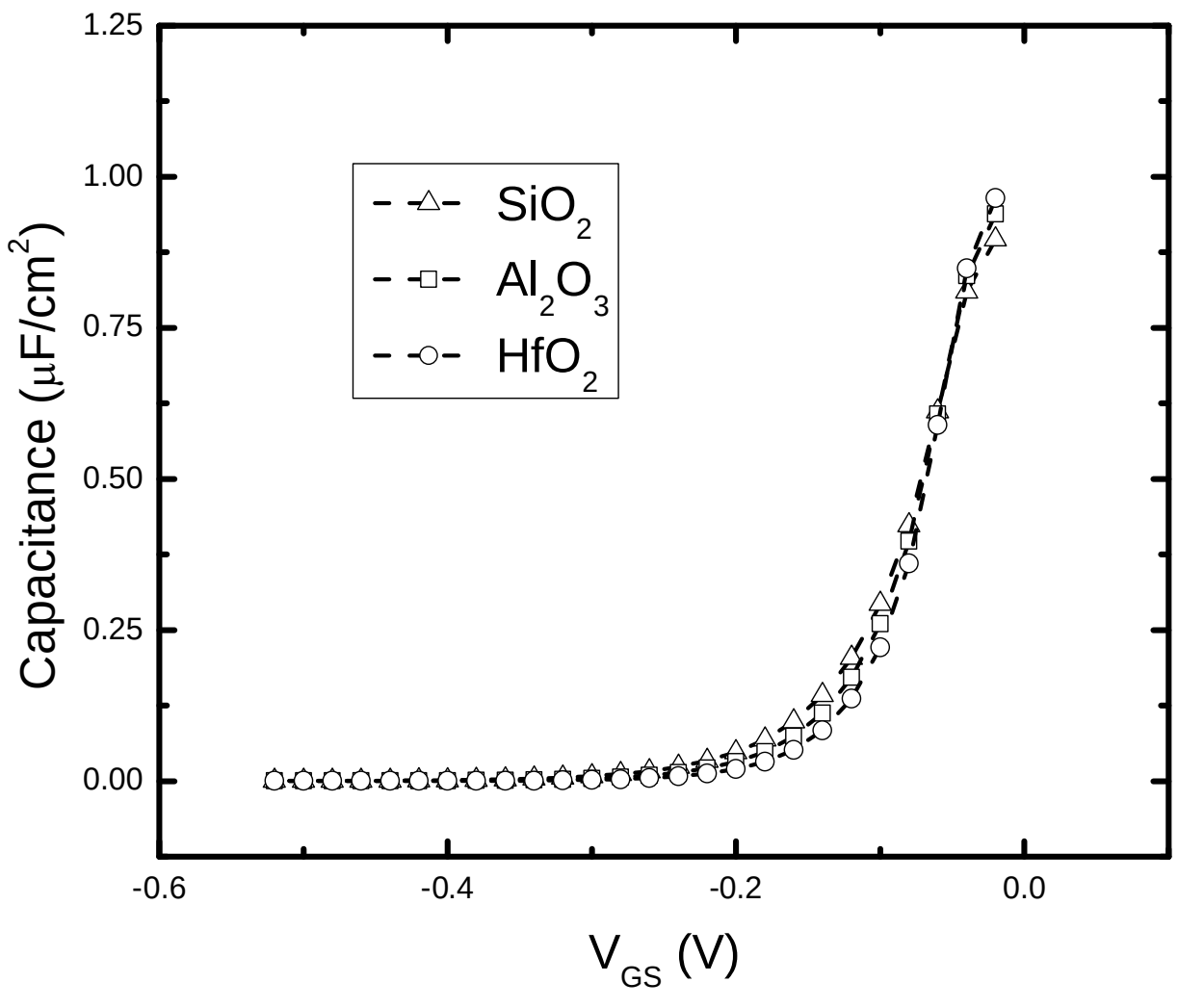

Figure 8. Gate capacitance as a function of gate voltage for different dielectric materials.

First eigen energy, channel sheet carrier density and gate capacitance for different oxide thickness is illustrated in Fig. 9, 10 and 11 respectively. It is observed at more negative voltages the capacitance is highest for the highest oxide thickness. But with the increase of gate voltage there is a cross over of capacitances and then capacitance becomes highest for the lowest oxide thickness. This result is evident from the increasing slope of the channel sheet carrier density for lowest gate oxide thickness after the cross over.

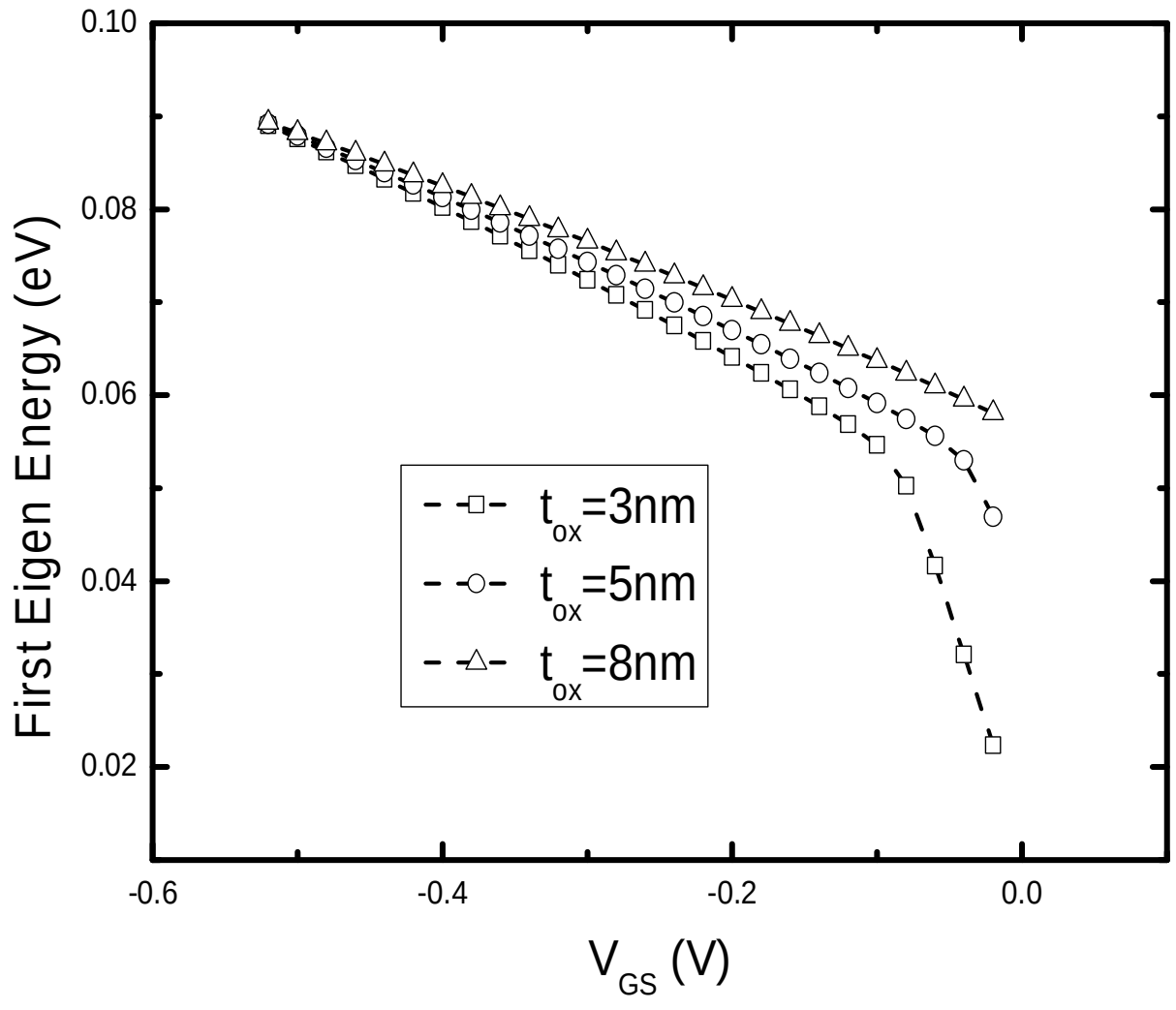

Figure 9. First eigen energy as a function of gate voltage for different oxide thickness with $Al_2O_3$ as dielectric and In percentage 0.60

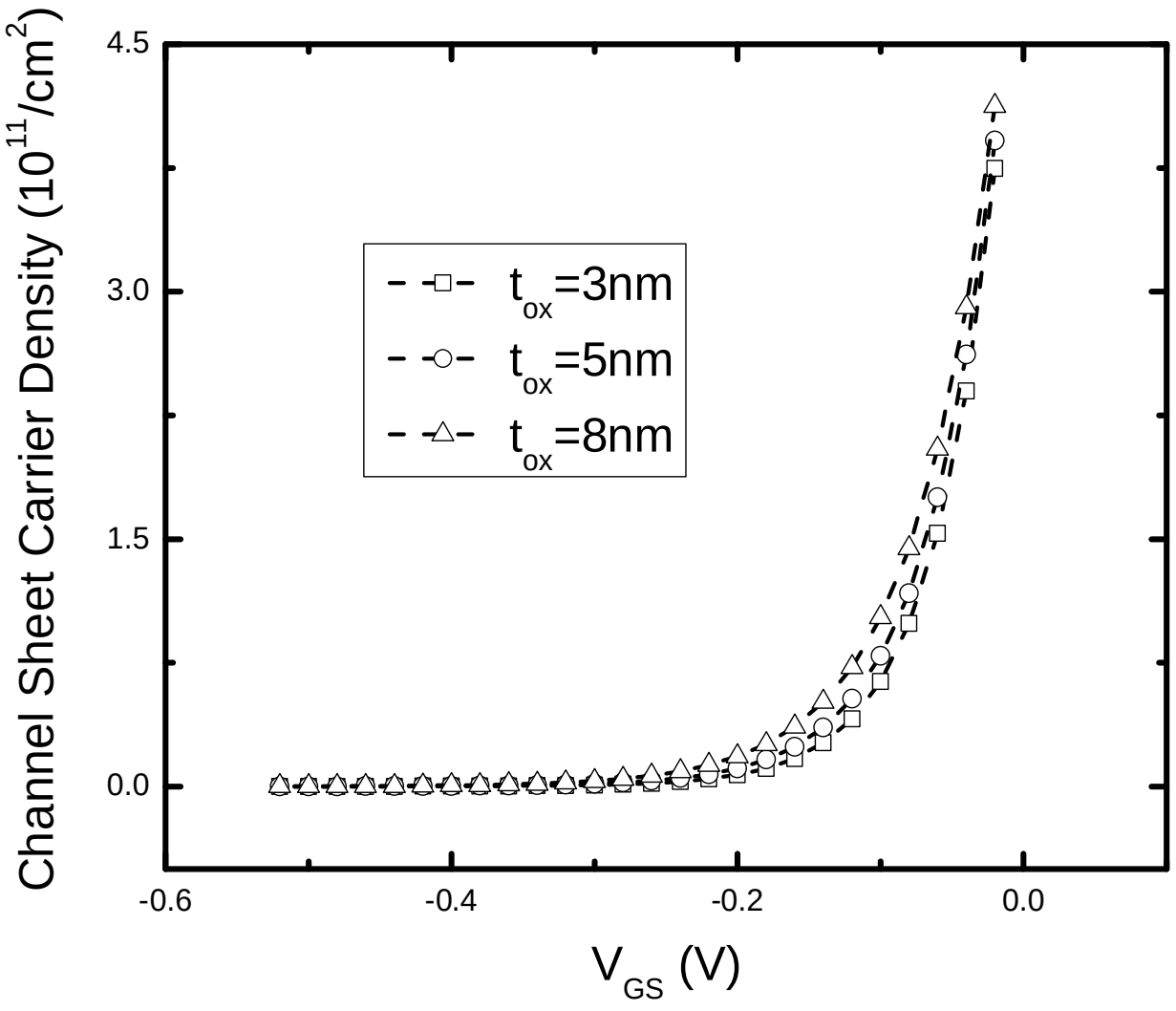

Figure 10. Channel sheet carrier density as a function of gate voltage for different oxide thickness.

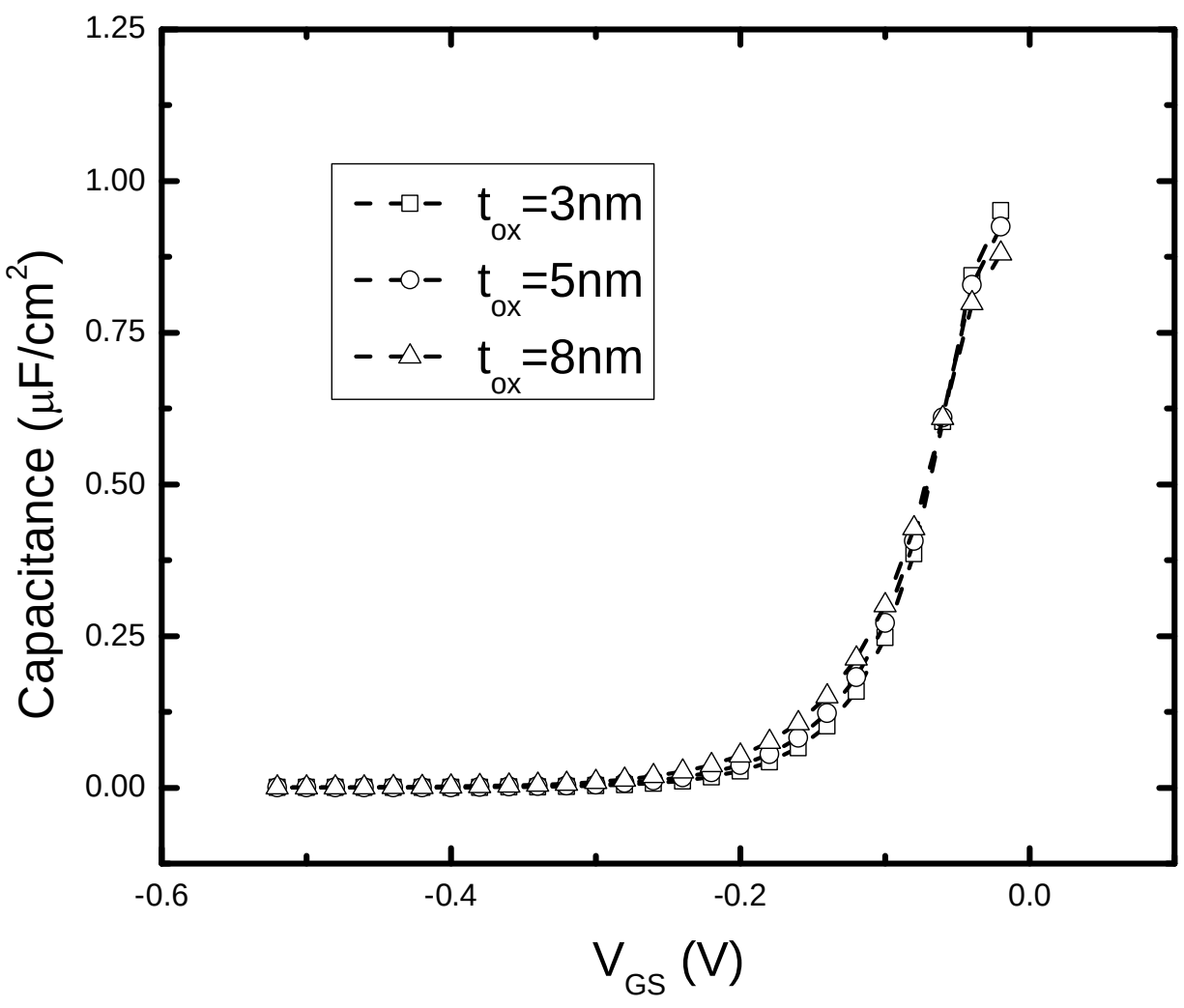

Figure 11. Gate capacitance as a function of gate voltage for different oxide thickness.

We also investigated the ballistic drain current for different In compositions in the top barrier region to see how strain affects drain current. This is illustrated in Fig. 12. We found that higher In composition i.e. higher tensile strain between channel and top barrier increases the drain current significantly.

## IV. Conclusions

Self consistent simulation for Buried Channel InGaAs/InAs QWFET has been demonstrated taking wave function penetration and other QM effects along with strain effect into account. A complete device characterization is done with a detailed analysis on the effect of parameter variation on the device performance. The ballistic drain current was also obtained to investigate the improvement of device performance. It is observed that inversion capacitance and ballistic current increases with the increase of Indium (In) mole fraction in barrier layer which yields better device performance.

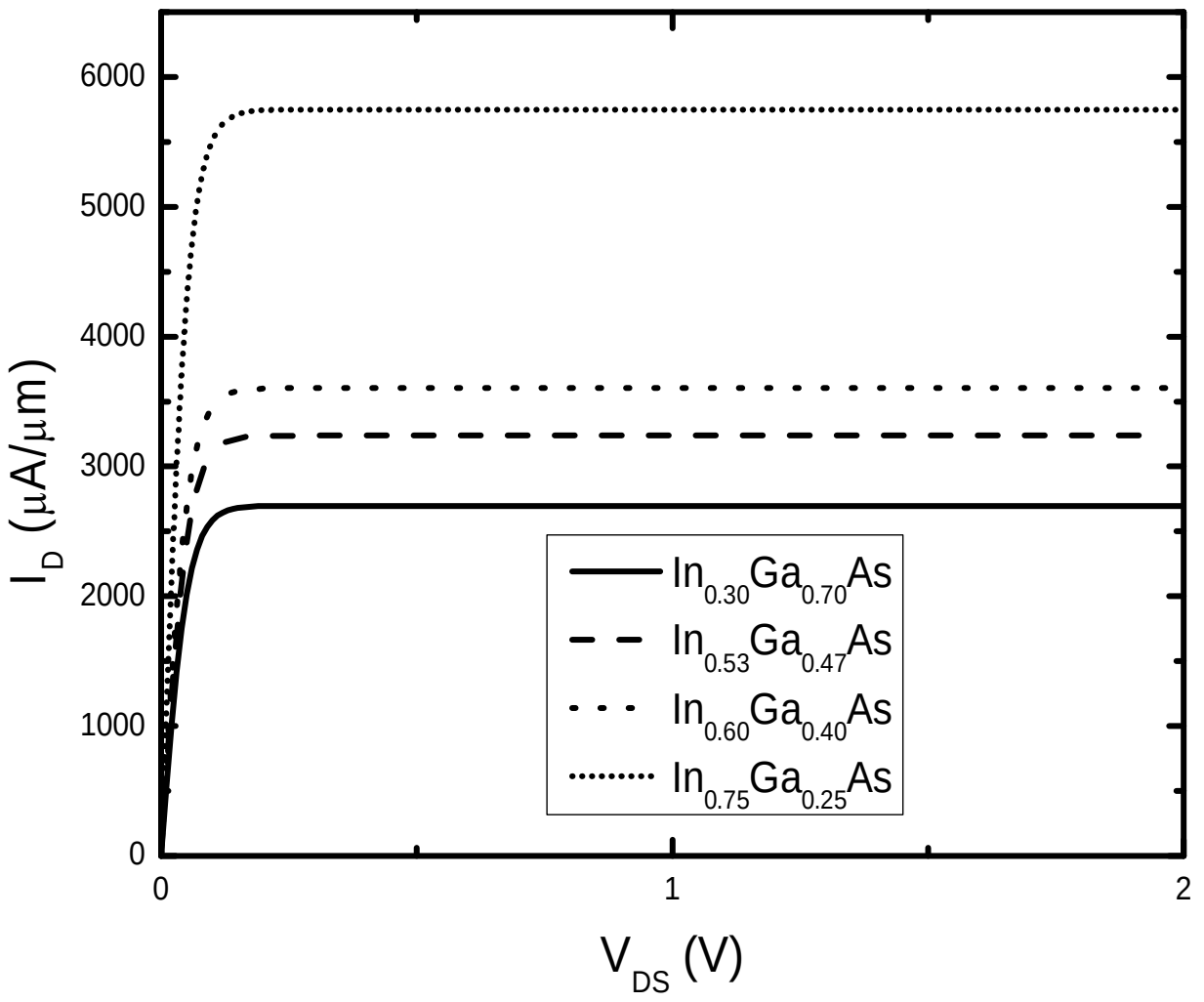

Figure 12. Ballistic drain current as a function of drain voltage for different In composition at top barrier for $L_G = 20$nm.

## Acknowledgment

The authors would like to thank Mr. Raisul Islam for his technical support and guidance in simulation procedure.